\documentclass[a4paper]{article}
\usepackage[german, english]{babel}
\usepackage{a4wide}
\usepackage{amsmath}
\usepackage{amssymb}
\usepackage{ifthen}
\usepackage{epsfig}
\newcounter{fig}   

\newcommand{\vphi}{\varphi}

\newcommand{\sqt}{\sqrt{3}}

\begin{document}

\title{\bf Excited Platonic Sphalerons\\ in the Presence of a Dilaton Field}
\vspace{1.5truecm}
\author{
{\bf Burkhard Kleihaus, Jutta Kunz and Kari Myklevoll}\\
Institut f\"ur  Physik, Universit\"at Oldenburg, Postfach 2503\\
D-26111 Oldenburg, Germany}

\vspace{1.5truecm}

\date{\today}

\maketitle
\vspace{1.0truecm}

\begin{abstract}
We construct sphaleron solutions with discrete symmetries
in Yang-Mills-Higgs theory coupled to a dilaton.
These platonic sphalerons are related to rational maps of degree $N$.
We demonstrate that, in the presence of a dilaton,
for a given rational map excited platonic sphalerons exist
beside the fundamental platonic sphalerons.
We focus on platonic sphaleron solutions with $N=4$,
which possess cubic symmetry,
and construct the two branches of their first excitations.
The energy density of these excited platonic sphalerons exhibits
a cube within a cube.
\end{abstract}
%\vfill\eject

\section{Introduction}

The non-trivial topology of configuration space of
SU(2) Yang-Mills-Higgs (YMH) theory
gives rise to a variety of static, finite energy solutions.
The simplest classical solutions possess spherical symmetry. 
In the presence of a triplet Higgs field 
these are the topologically stable 't Hooft-Polyakov monopoles 
\cite{thooft},
whereas in the presence of a doublet Higgs field 
these are the unstable sphalerons \cite{km,bi,yaffe}.

Multimonopoles and multisphalerons represent
YMH solutions with less symmetry.
Characterizing multimonopoles by their topological charge $N$, $N \ge 2$,
and multisphalerons by their Chern-Simons number $N/2$,
these solutions possess at most axial symmetry
\cite{monoax,kksph}.

For $N \ge 3$ multimonopole and multisphaleron solutions 
with no rotational symmetry appear \cite{monopla,kkm}.
The symmetries of these solutions are only discrete,
and can be identified with the symmetries of platonic solids or crystals.
We therefore refer to them as platonic monopoles and 
platonic sphalerons.
Related to certain rational maps of degree $N$ \cite{ratmap},
they have many properties in common.

In flat space the static classical solutions associated with
a certain rational map of degree $N$ appear to be unique.
This has been shown explicitly 
for the 't Hooft-Polyakov monopole \cite{maison}
and the Klinkhamer-Manton sphaleron \cite{yaffe},
which do not possess radial excitations\footnote{
An exception are bisphalerons, present for very large
Higgs mass \cite{bi,yaffe}.}.

In curved space, in contrast, the absence of radially
excited monopole and sphaleron solutions is no longer true.
Indeed, when gravity is coupled, both gravitating
monopoles and gravitating sphalerons possess a whole
sequence of radial excitations \cite{gmono,greene}.
These radially excitated 
Einstein-Yang-Mills-Higgs (EYMH) solutions
are related to the Bartnik-McKinnon solutions
of pure Einstein-Yang-Mills (EYM) theory \cite{bm,vg}.

The coupling of gravity to Yang-Mills or YMH theory
is known to have a very similar effect 
concerning the existence and the properties of 
classical spherically symmetric solutions 
as the coupling of a scalar dilaton \cite{lav,gmono,forgacs}.
Dilatonic monopoles possess
a sequence of radial excitations
analogous to gravitating monopoles \cite{forgacs},
and dilatonic sphalerons possess 
a sequence of radial excitations
analogous to gravitating sphalerons, as demonstrated below.

The existence of excited spherically symmetric solutions 
as well as of excited axially symmetric solutions in the presence
of gravity \cite{kk} 
and analogously in the presence of a dilaton \cite{kk-dil}
suggests, that also excited platonic solutions might exist.
We here demonstrate, that this is indeed the case,
by explicitly constructing excited platonic sphalerons
in the presence of a dilaton.
These excited platonic sphalerons represent the first examples 
of excited platonic solutions.

We focus on platonic Yang-Mills-Higgs-dilaton (YMHD) sphalerons,
based on a rational map of degree $N=4$,
which possess cubic symmetry \cite{kkm,kkm-dil}.
Like the fundamental platonic YMHD sphalerons \cite{kkm-dil},
excited platonic sphalerons should form two branches of solutions,
merging at a maximal value of the coupling constant.
The maximal value of the coupling constant
is expected to decrease with increasing excitation.
We here explicitly construct the two branches of the first excitations
of the platonic YMHD sphalerons.
We also construct excited spherically and axially symmetric YMHD sphalerons.

We review YMHD theory in section 2. 
We present the Ans\"atze and the boundary conditions for axially symmetric
and platonic sphalerons in section 3,
and discuss our numerical results in section 4.

\section{Yang-Mills-Higgs-Dilaton Theory}

We consider YMHD theory with Lagrangian
\begin{equation}
{\cal L} = -\frac{1}{2} \partial_\mu \phi \partial^\mu \phi
-\frac{1}{2} e^{2\kappa \phi} {\rm Tr} (F_{\mu\nu} F^{\mu\nu})
- (D_\mu \Phi)^\dagger (D^\mu \Phi) 
- \lambda e^{-2\kappa \phi} (\Phi^\dagger\Phi - \frac{v^2}{2} )^2
\  
\label{lag1}
\ , \end{equation}
SU(2) field strength tensor
\begin{equation}
F_{\mu\nu}=\partial_\mu V_\nu-\partial_\nu V_\mu
            + i g [V_\mu , V_\nu ]
\ , \end{equation}
SU(2) gauge potential $V_\mu = V_\mu^a \tau_a/2$,
covariant derivative of the Higgs doublet $\Phi$
\begin{equation}
D_{\mu} \Phi = \partial_{\mu}\Phi+i g V_\mu  \Phi
\ , \end{equation}
and dilaton field $\phi$,
where $g$ and $\kappa$ denote the gauge and dilaton coupling constants,
respectively,
$\lambda$ denotes the strength of the Higgs self-interaction, and
$v$ the vacuum expectation value of the Higgs field.

The Lagrangian (\ref{lag1}) is invariant under local SU(2)
gauge transformations $U$,
\begin{eqnarray}
V_\mu &\longrightarrow & U V_\mu U^\dagger
+ \frac{i}{g} \partial_\mu U  U^\dagger \ ,
\nonumber\\
\Phi  &\longrightarrow & U \Phi \ .
\nonumber
\end{eqnarray}
The gauge symmetry is spontaneously broken 
due to the non-vanishing vacuum expectation
value of the Higgs field
\begin{equation}
    \langle \Phi \rangle = \frac{v}{\sqrt2}
    \left( \begin{array}{c} 0\\1  \end{array} \right)   
\ . \end{equation}
In the limit of vanishing dilaton coupling constant,
the model corresponds to the bosonic sector of
Weinberg-Salam theory for vanishing Weinberg angle.

In the following we consider only static finite energy solutions, with
$V_0=0$, $V_i = V_i(\vec{r}\,)$, $i=1,2,3$, $\Phi = \Phi(\vec{r}\,)$, 
$\phi = \phi(\vec{r}\,)$.
The energy of such solutions is given by 
\begin{equation}
E = \int\left( \frac{1}{2} \partial_i \phi \partial^i \phi
+\frac{1}{2} e^{2\kappa \phi} {\rm Tr} (F_{ij} F^{ij})
+ (D_i \Phi)^\dagger (D^i \Phi) 
+ \lambda e^{-2\kappa \phi} (\Phi^\dagger \Phi - \frac{v^2}{2})^2  
\right) d^3r \ .
\label{energy}
\end{equation}
The energy $E$ is related to
the dilaton charge $D$,
\begin{equation}
D = \frac{1}{4\pi} \int_{S_2} \vec{\nabla}\phi \cdot d\vec{S} \ ,
\end{equation}
by the simple relation
\begin{equation}
4 \pi D = \kappa E \ .
\label{EDrel}
\end{equation}
This can be seen by integrating the dilaton equation and using the identity
\begin{equation}
0=\int\left( -\frac{1}{2} \partial_i \phi \partial^i \phi
+\frac{1}{2} e^{2\kappa \phi} {\rm Tr} (F_{ij} F^{ij})
+ (D_i \Phi)^\dagger (D^i \Phi) 
+ 3 \lambda e^{-2\kappa \phi} (\Phi^\dagger \Phi - \frac{v^2}{2} )^2 
\right) d^3r
\ , \label{derrick} 
\end{equation}
which follows from a Derrick like argument
for solutions of the field equations.

The Chern-Simons charge $N_{\rm CS}$ of the solutions is given by
\begin{equation}
 N_{\rm CS} = \int d^3r K^0 \ ,
\label{Q}
\end{equation}
with
\begin{equation}
 K^\mu=\frac{g^2}{16\pi^2}\varepsilon^{\mu\nu\rho\sigma} {\rm Tr}(
 F_{\nu\rho}V_\sigma
 + \frac{2}{3} i g V_\nu V_\rho V_\sigma )
\ . \end{equation}
For the identification of the baryon number 
$Q_{\rm B}=N_{\rm CS}$ \cite{km,kkm}, 
the Chern-Simons number
$N_{\rm CS}$ has to be evaluated in a gauge, where asymptotically
\begin{equation}
V_\mu \to \frac{i}{g} \partial_\mu \hat{U} \hat{U}^\dagger \ , \ \ \
\hat{U}(\infty) = 1 \ .
\end{equation}

\section{Sphaleron Solutions}

We here present the Ans\"atze and the boundary conditions for 
axially symmetric and platonic sphalerons.
The excited sphalerons fall into two classes,
distinguished by their boundary conditions and their
Chern-Simons charges.

We parametrize the Higgs field as
\begin{equation}
\Phi = (\Phi_0 1\hspace{-0.28cm}\perp + i \Phi_a \tau_a)\frac{v}{\sqrt2}
    \left( \begin{array}{c} 0\\1  \end{array} \right) \ ,
\end{equation}
and impose at infinity the boundary conditions
\begin{equation}
\Phi_0 = 0 \ , \ \ \
\Phi_a  \tau_a=   \vec {n}_R \cdot {\vec \tau} =: \tau_R\ ,
\label{bcHiggs}
\end{equation}
where $\tau_a$ are the Cartesian Pauli matrices
and $\vec {n}_R$ is a unit vector to be specified.
The boundary conditions for the gauge field at infinity
are then obtained from the requirement $D_i \Phi =0$. 

When the Higgs field is angle-dependent at infinity
this yields for the gauge field
\begin{equation}
V_i = \frac{i}{g} (\partial_i \tau_R )\tau_R
\ ,
\label{bcA}
\end{equation}
i.e., the gauge field tends to a pure gauge at infinity,
$V_i = \frac{i}{g} (\partial_i U_\infty) U_\infty^\dagger $,
with $U_\infty=i \tau_R$. 
The baryon number of the sphalerons is then obtained from
Eq.~(\ref{Q}), after performing a gauge transformation with \cite{kkm}
\begin{equation}
U = \exp(-i \Omega(x,y,z) \tau_R) \ ,
\label{gtu} \end{equation}
where $\Omega$ tends to $\pi/2$ at infinity and vanishes at
the origin.

In contrast, 
when the Higgs field tends to a constant isospinor, 
e.g.~when $\Phi_a  \tau_a=\tau_z$,
the gauge field vanishes identically at infinity,
and the baryon number then vanishes as well.

For the dilaton field we require that it vanishes at infinity,
$\phi_{\infty}=0$,
since any finite value of the dilaton field at infinity
can always be transformed to zero via
$\phi \rightarrow \phi - \phi_\infty$,
$r \rightarrow r e^{-\kappa \phi_\infty} $.

\subsection{Axially Symmetric Sphalerons}

The Ansatz for the axially symmetric YMHD sphalerons 
corresponds to the Ansatz employed in Weinberg-Salam theory \cite{kksph}
(at vanishing Weinberg angle \cite{kkb}),
\begin{equation}
V_i dx^i = \left(\frac{H_1}{r} dr + (1-H_2) d\theta\right)
           \frac{\tau^{(n)}_\vphi}{2g}
          -n\sin\theta\left(H_3 \frac{\tau^{(n)}_r}{2g} 
	  + (1-H_4)\frac{\tau^{(n)}_\theta}{2g}\right) d\vphi
	   \ , \ \ \ V_0=0 \ ,
\label{a_axsym}
\end{equation}	  
and
\begin{equation}
\Phi = i( \Phi_1 \tau^{(n)}_r + \Phi_2 \tau^{(n)}_\theta )\frac{v}{\sqrt2}
    \left( \begin{array}{c} 0\\1  \end{array} \right) \ ,
\end{equation}
supplemented by the dilaton function $\phi$,
where
\begin{eqnarray}	  
\tau^{(n)}_r & = & \sin\theta (\cos n\vphi \tau_x + \sin n\vphi \tau_y) 
           + \cos\theta \tau_z \ , \ \ 
\nonumber \\	   
\tau^{(n)}_\theta & = & \cos\theta (\cos n\vphi \tau_x + \sin n\vphi \tau_y) 
           - \sin\theta \tau_z \ , \ \ 
\nonumber \\	   
\tau^{(n)}_\vphi & = & (-\sin n\vphi \tau_x + \cos n\vphi \tau_y) 
\ , \ \ \nonumber 
\end{eqnarray}	  
$\tau_x$, $\tau_y$ and $\tau_z$ denote the Pauli matrices,
and $n$ is an integer.
For $n=1$ and $\kappa=0$ the Ansatz yields the 
spherically symmetric Klinkhamer-Manton sphaleron \cite{km}.
For $n>1$,
the functions $H_1$--$H_4$, $\Phi_1$, $\Phi_2$ and 
$\phi$ depend on $r$ and $\theta$, only. 
With this Ansatz the full set of field equations reduces to a system 
of seven coupled partial differential equations in the independent variables 
$r$ and $\theta$. A residual U(1) gauge degree of freedom is 
fixed by the condition $r\partial_r H_1 - \partial_\theta H_2=0$ \cite{kksph}.

In the presence of gravity or of a dilaton,
for each integer $n$, a whole sequence of 
regular axially symmetric sphaleron solutions exists,
labelled by the node number $k$ of
the gauge field functions $H_2$ and $H_4$ \cite{bm,kk,greene}.
The sphaleron solutions then
fall into two classes, odd-$k$ sphalerons
with the fundamental sphaleron ($k=1$) as their lowest mass member,
and even-$k$ sphalerons, with the first excited
sphaleron ($k=2$) as their lowest mass member.

The boundary conditions differ for the two classes of solutions
\cite{bm,kk,greene}.
Regularity at the origin requires the boundary conditions
\begin{equation}
H_1=H_3=0 \ , \ H_2=H_4=1 \ , \ \partial_r \phi =0 \ , \ \ \ \ {\rm all\ } k \ ,
\end{equation}
\begin{equation}
\Phi_1=\Phi_2=0 \ , \ \ \ \ {\rm odd\ } k \ ,
\end{equation}
\begin{equation}
\sin \theta\, \Phi_1 + \cos \theta\, \Phi_2 =0 \ , \
\cos \theta\, \Phi_1 - \sin \theta\, \Phi_2 = \Phi_c \ , \ \ \ \ {\rm even\ } k \ ,
\end{equation}
where $\Phi_c$ is a finite constant \cite{greene}. 
Thus the Higgs field vanishes at the origin only for odd-$k$ sphalerons.
In contrast the gauge field vanishes at the origin for all sphalerons
(in the chosen gauge).
Since $\Phi_c$ is unknown, however, we employ the boundary conditions
\begin{equation}
\sin \theta\, \Phi_1 + \cos \theta\, \Phi_2 =0 \ , \
\partial_r (\cos\theta\, \Phi_1 - \sin\theta\, \Phi_2) = 0 \ , \ \ \ \ {\rm even\ } k \ .
\end{equation}

Regularity on the $z$-axis requires the boundary conditions
\begin{equation}      
H_1=H_3=\Phi_2=0 \ , \
\partial_\theta\, H_2=\partial_\theta\, H_4=\partial_\theta\, \Phi_1
=\partial_\theta\, \phi=0 \ , \ \ \ \ {\rm all\ } k \ .
\end{equation}
% distinction not necessary here
%\begin{equation}
%\partial_\theta\, \Phi_1=1 \ , \ \Phi_2=0 \ , \ \ \ \ {\rm odd\ } k \ ,
%\end{equation}
%\begin{equation}
%\sin \theta\, \Phi_1 + \cos \theta\, \Phi_2 =0 \ , \
%\cos \theta\, \partial_\theta\, \Phi_1 - \sin \theta\, \partial_\theta\, \Phi_2 = 0 \ , 
%\ \ \ \ {\rm even\ } k \ .
%\end{equation}

The requirement of finite energy implies 
at infinity the boundary conditions
\begin{equation}
H_1=H_3=0 \ , \ H_2=H_4=(-1)^k \ , \ \phi =\phi_\infty \ ,  
\ \ \ \ {\rm all\ } k \ ,
\end{equation}
\begin{equation}
\Phi_1=1 \ , \ \Phi_2=0 \ , \ \ \ \ {\rm odd\ } k \ ,
\end{equation}
\begin{equation}
\sin \theta\, \Phi_1 + \cos \theta\, \Phi_2 =0 \ , \
\cos \theta\, \Phi_1 - \sin \theta\, \Phi_2 = 1 \ , \ \ \ \ {\rm even\ } k \ .
\end{equation}
Thus for odd-$k$ sphalerons,
the Higgs field is angle-dependent at infinity,
$\Phi_a  \tau_a=\tau_r^{(n)}$,
and the gauge field then tends to a pure gauge configuration
with $U_\infty=i \tau_r^{(n)}$.
In contrast, for even-$k$ sphalerons,
the Higgs field tends to a constant isospinor,
$\Phi_a  \tau_a=\tau_z$, and thus the gauge field vanishes at infinity.

The non-gauge transformed Chern-Simons density $K^0$
vanishes identically for the spherically and axially symmetric sphalerons,
because of the ansatz of the gauge potential.
After performing the gauge transformation (\ref{gtu}),
%($\tau_R = \vec {n}_R \cdot {\vec \tau}$), 
we obtain \cite{kb,kksph,kkm}
\begin{equation}
N_{\rm CS}=n/2  \ , \ \ \ {\rm odd\ } k \ , \ \ \ 
N_{\rm CS}=0    \ , \ \ \ {\rm even\ } k \ .
\end{equation}
Thus for a given integer $n$,
odd-$k$ sphalerons, and in particular fundamental sphalerons,
carry baryon number $Q_{\rm B}=n/2$,
whereas even-$k$ sphalerons have vanishing baryon number.

\subsection{Platonic Sphalerons}

To obtain YMHD solutions with discrete symmetry
we make use of rational maps,
i.e.~holomorphic functions from $S^2\mapsto S^2$ \cite{ratmap}.
Treating each $S^2$ as a Riemann sphere, the first having coordinate 
$\xi$,
a rational map of degree $N$ is a function $R:S^2\mapsto S^2$ where
\begin{equation}
R(\xi)=\frac{p(\xi)}{q(\xi)} 
\ , \label{rat} \end{equation}
and $p$ and $q$ are polynomials of degree at most $N$, where at least
one of $p$ and $q$ must have degree precisely $N$, and $p$ and $q$
must have no common factors \cite{ratmap}.

We recall that via stereographic projection, the complex coordinate $\xi$
on a sphere can be identified with conventional polar coordinates by
$\xi=\tan(\theta/2)e^{i\varphi}$ \cite{ratmap}.
Thus the point $\xi$ corresponds to the unit vector
\begin{equation}
\vec {n}_\xi=\frac{1}{1+\vert \xi \vert^2}
(2\Re(\xi), 2\Im(\xi),1-\vert \xi \vert^2)
\ , \label{unit1} \end{equation}
and the value of the rational map $R(\xi)$ 
is associated with the unit vector
\begin{equation}
\vec {n}_R=\frac{1}{1+\vert R \vert^2}
(2\Re(R), 2\Im(R),1-\vert R\vert^2) \ .
\label{unit2}
\end{equation}

We here consider platonic YMHD solutions obtained from the map $R_4$,
\begin{equation}
R_4(\xi)=\frac{\xi^4+2\sqrt{3}i\xi^2+1}{\xi^4-2\sqrt{3}i\xi^2+1}  \ \ \ . \
\label{map2} \end{equation}
Note, that the map $R_4(\xi)= \xi^4$ yields
the axially symmetric sphalerons for $n=4$ in a different gauge.

The rational map Eq.~(\ref{map2}) leads to the unit vector (\ref{unit2})
\begin{equation}
\vec {n}_{R_4}= \left(-\frac{(r^2-z^2)^2-2 z^2 r^2+2 x^2 y^2}
                        {{\cal N}_4} \ ,
                   \frac{\sqt (r^2+z^2)(x^2-y^2)}
                        {{\cal N}_4} \ ,
                   \frac{4 \sqt r x y z}
                        {{\cal N}_4}
             \right) \ ,
\label{unit4}
\end{equation}
with ${\cal N}_4 =2(x^4+x^2 y^2+x^2 z^2+y^4+y^2 z^2+z^4)$.

The boundary conditions for platonic sphalerons are,
in principle, only needed at infinity. 
Specifying the unit vector $\vec {n}_R$ thus completely
determines the boundary conditions for the Higgs field (\ref{bcHiggs})
and for the gauge field (\ref{bcA}).
For fundamental $N=4$ platonic sphalerons with cubic symmetry
$\vec {n}_R= \vec {n}_{R_4}$.
Thus at infinity the Higgs field
and the gauge field exhibit the angle-dependence of the rational map, 
since $\Phi_a  \tau_a=\tau_{R_4}$ and $U_\infty=i\tau_{R_4}$.
In contrast, for the first excited $N=4$ platonic sphalerons
an adequate choice for the Higgs field
at infinity is again $\Phi_a  \tau_a=\tau_z$,
just as for the first excited rotationally symmetric sphalerons.
Consequently the gauge field vanishes at infinity,
and the discrete symmetry of the first excited platonic sphalerons
does not enter via the boundary conditions at infinity.

For platonic sphalerons the non-gauge transformed
Chern-Simons density $K^0$ is non-trivial,
but as checked numerically it does not contribute 
to their baryon number \cite{kkm}.
Thus the only contribution arises again from the gauge transformation $U$,
Eq.~(\ref{gtu}), yielding the Chern-Simons number
\begin{equation}
N_{\rm CS} = N/2 \ \ \ {\rm fundamental} \ , \ \ \
N_{\rm CS} = 0 \ \ \ {\rm first\ excitation} \ .
\end{equation}

The general set of field equations involves
the dilaton function $\phi(x,y,z)$,
three functions for the Higgs field, $\Phi_a(x,y,z)$,
and nine functions for the gauge field, $V_i^a(x,y,z)$,
while the remaining Higgs and gauge field functions vanish
consistently, $\Phi_0=V_0^a=0$.
Subject to the boundary conditions at infinity,
and to the gauge condition
\begin{equation}
\partial_i V^i =0
\ , \label{gaugecond} \end{equation}
one can then solve this set of field equations 
in the full space $\mathbb{R}^3$.

To obtain better numerical accuracy, however,
one can make use of the discrete symmetries of the platonic sphalerons,
to restrict the region of numerical integration.
After specifying the reflection symmetries of the fields
with respect to the $xy$-, $xz$- and $yz$-plane,
it is then sufficient to solve the field equations for $x\ge 0$ and 
$y\ge 0$ and $z\ge 0$ only. 

The boundary conditions at infinity must then be supplemented
by conditions at the other boundaries of the integration region.
These boundaries consist of the origin, the $xy$-, $xz$- and $yz$-plane,
and the positive $z$-axis.
(Together with the boundary at infinity these boundaries
are implemented numerically as the boundaries of a cube \cite{fidisol}.)
At the origin ${r}=0$, 
the Higgs field vanishes for the fundamental platonic sphalerons,
while it is finite for the first excited platonic sphalerons,
with boundary conditions $\Phi_1=\Phi_2=0$ and $\partial_r \Phi_3=0$.
The gauge field always vanishes at the origin,
and the dilaton field satisfies $\partial_{{r}} \phi = 0$.

The boundary conditions in the $xz$-plane ($\vphi=0$),  
the $yz$-plane ($\vphi=\pi/2$), on the $z$-axis 
($\theta=0$) and in the $xy$-plane ($\theta=\pi/2$)
follow from the reflection symmetries of the fields.
%For the rational map Eq.~(\ref{map2}) with unit vector (\ref{unit4})
%$n_R^1$ and $n_R^2$ are even, while $n_R^3$ is odd in $x$, $y$, $z$.
In particular, we suppose that the rational map $R_4$ determines
the reflection symmetries of the fields.
Thus we impose, that
for the fundamental platonic sphalerons
the Higgs field $\Phi$ and the
gauge field $V_i$ possess the same reflection symmetries 
as $\tau_{R_4}$ and $\frac{i}{g} [\partial_i \tau_{R_4} ,\tau_{R_4}]$, 
respectively\footnote{
When the platonic sphalerons are calculated numerically 
in the full space $\mathbb{R}^3$ 
they satisfy the respective reflection symmetries.}.
For the first excited platonic sphalerons, however,
the boundary conditions of the Higgs field in the planes 
must also be compatible
with the boundary conditions at the origin and at infinity,
i.e.,~$\Phi_a \tau_a \sim \tau_z$.
The consistent set of Higgs boundary conditions then follows
from the symmetry requirements of the energy density,
while the boundary conditions for the gauge field remain unchanged.
For the Higgs and spherical gauge field components
the proper set of boundary conditions is given in Table 1.
For the dilaton field the normal derivative must vanish in 
the $xy$-, $xz$- and $yz$-plane, 
and on the $z$-axis $\partial_\theta \phi =0$. 

\begin{table}
\begin{tabular}{|c|c|c|}
\hline
$k$ & $xz$-plane ($\varphi=0$) & $yz$-plane ($\varphi=\pi/2$)
\\ \hline
1, 2 & \begin{tabular}{ccc}
       $V_r^1=0$ \ , &   $V_r^2=0$ \ ,  &  $\partial_\vphi V_r^3=0$ \ , \\
       $V_\theta^1=0$ \ , &   $V_\theta^2=0$ \ ,  &   
       $\partial_\vphi V_\theta^3=0$ \ , \\
       $\partial_\vphi V_\vphi^1=0$ \ ,  &  $\partial_\vphi V_\vphi^2=0$ \ , &   
       $V_\vphi^3=0$ \ ,
     \end{tabular} 
   & \begin{tabular}{ccc}
       $V_r^1=0$ \ ,  &   $V_r^2=0 $\ ,   &   $\partial_\vphi V_r^3=0 $\ , \\
       $V_\theta^1=0$ \ ,   &   $V_\theta^2=0$ \ ,   &  $\partial_\vphi  V_\theta^3=0$ \ , \\
       $\partial_\vphi V_\vphi^1=0$ \ ,   &   $\partial_\vphi V_\vphi^2=0$ \ ,   &  
       $V_\vphi^3=0$ \ ,
     \end{tabular} 
 \\ \hline
1 & \begin{tabular}{ccc}
       $\partial_\vphi \Phi_1=0$ \ , &  $\partial_\vphi \Phi_2=0$ \ ,  &  $\Phi_3=0$ \ ,
     \end{tabular} 
   & \begin{tabular}{ccc}
       $\partial_\vphi \Phi_1=0$ \ ,  &  $\partial_\vphi \Phi_2=0$ \ ,  &   $\Phi_3=0$ \ , 
     \end{tabular} 
 \\ \hline
2 & \begin{tabular}{ccc}
       $\Phi_1=0$ \ , &  $\Phi_2=0$ \ ,  &  $\partial_\vphi \Phi_3=0$ \ ,
     \end{tabular} 
   & \begin{tabular}{ccc}
       $\Phi_1=0$ \ ,  &  $\Phi_2=0$ \ ,  &   $\partial_\vphi \Phi_3=0$ \ , 
     \end{tabular} 
 \\ \hline
    & $z$-axis ($\theta=0$) & $xy$-plane ($\theta=\pi/2$) 
\\ \hline  
1, 2 & \begin{tabular}{ccc}
     $V_r^1=0$ \ , & $V_r^2=0$ \ , & $\partial_\theta V_r^3=0$ \ , \\
     $V_\theta^1=0$ \ , & $V_\theta^2=0$ \ , & $V_\theta^3=0$ \ , \\
     $V_\vphi^1=0$ \ , & $V_\vphi^2=0$ \ , & $V_\vphi^3=0$ \ ,
     \end{tabular} 
  & \begin{tabular}{ccc}
     $V_r^1=0$ \ , & $V_r^2=0$ \ , & $\partial_\theta V_r^3=0$ \ , \\
     $\partial_\theta V_\theta^1=0$ \ , & $\partial_\theta V_\theta^2=0$ \ , & 
     $V_\theta^3=0$ \ , \\
     $V_\vphi^1=0$ \ , & $V_\vphi^2=0$ \ , & $\partial_\theta V_\vphi^3=0$ \ ,
     \end{tabular} 
 \\ \hline
1 & \begin{tabular}{ccc}
       $\partial_\theta \Phi_1=0$ \ , &  $\partial_\theta \Phi_2=0$ \ ,  &  $\Phi_3=0$ \ ,
     \end{tabular} 
   & \begin{tabular}{ccc}
       $\partial_\theta \Phi_1=0$ \ ,  &  $\partial_\theta \Phi_2=0$ \ ,  &   $\Phi_3=0$ \ , 
     \end{tabular} 
 \\ \hline
2 & \begin{tabular}{ccc}
       $\Phi_1=0$ \ , &  $\Phi_2=0$ \ ,  &  $\partial_\theta \Phi_3=0$ \ ,
     \end{tabular} 
   & \begin{tabular}{ccc}
       $\Phi_1=0$ \ ,  &  $\Phi_2=0$ \ ,  &   $\partial_\theta \Phi_3=0$ \ , 
     \end{tabular} 
\\ \hline
\end{tabular} 
\caption{ \small
The boundary conditions 
in the $xy-$, $xz-$, and $yz$-plane, and on the $z$-axis
are given for the gauge and Higgs field components
of the fundamental platonic sphalerons ($k=1$) and 
of the first excited platonic sphalerons ($k=2$).}
\end{table}

For the numerical construction of platonic sphalerons
an adequate initial guess is mandatory.
For the fundamental cubic sphalerons the initial guess is obtained
by parametrizing the Higgs and gauge fields according to
\begin{equation}
\Phi = i h(r) \tau_{R_4} \frac{v}{\sqrt2}
    \left( \begin{array}{c} 0\\1  \end{array} \right) \ , \ \ \
V_i =
\frac{1-f(r)}{2} \frac{i}{g} (\partial_i \tau_{R_4} )\tau_{R_4}
\ , \label{guess1}
\end{equation}
and the dilaton field by $\phi(r)$. 
This simple ansatz with the three radial functions
$h(r)$, $f(r)$, and $\phi(r)$
is inserted into the action,
the angular dependence is integrated,
and a set of variational equations is obtained,
similar to the set of spherically symmetric ($n=1$) sphaleron equations.
These ordinary differential equations are solved numerically,
subject to the set of boundary conditions of the spherical ($n=1$) sphalerons
(in particular $f(0)=1$, $f(\infty)=-1$).
The initial guess is then given by (\ref{guess1})
with these numerically determined functions.

For the first excited cubic sphalerons the initial guess
is obtained analogously, only the procedure now involves 
the parametrization
\begin{equation}
\Phi = i h(r) \tau_z \frac{v}{\sqrt2}
    \left( \begin{array}{c} 0\\1  \end{array} \right) \ , \ \ \
V_i =
\frac{1-f(r)}{2} \frac{i}{g} (\partial_i \tau_{R_4} )\tau_{R_4} 
\ , \label{guess2}
\end{equation}
consistent with the different set of boundary conditions
(in particular $f(0)=1$, $f(\infty)=1$).

\section{\bf Numerical results}

For convenience
we rescale the coordinates $\vec{r} \to \vec{r}/g v$,
the gauge potential $V_i \to v V_i$, 
and the dilaton field $\phi \to \phi/\kappa$,
and introduce the dimensionless coupling constant
\begin{equation}
\alpha = \kappa v \ .
%\beta = M_{\rm H}/M_{\rm W} \ .
\label{cc}
\end{equation}
We further rescale the dilaton charge $D \to D/\kappa g v$ and
the energy $E \to E 4 \pi v /g$,
to obtain the scaled dilaton charge--energy relation (\ref{EDrel})
\begin{equation}
D = \alpha^2 E \ .
\label{EDreld}
\end{equation}

The numerical solutions are constructed with help of the 
software package FIDISOL \cite{fidisol} based on the Newton-Raphson 
algorithm. 
The solutions are obtained in spherical coordinates $r$, $\theta$, $\vphi$.
To map the infinite range of the radial variable $r$ to the finite 
interval $[0,1]$ we introduce the compactified variable $\bar{r}=r/(1+r)$.
Typical grids contain $100\times 30$ points for 
the axially symmetric solutions and $80 \times 25 \times 25$ points
for the platonic solutions.
The estimated relative errors are on the order of $0.01$\% 
for the axially symmetric sphalerons, and 
$1$\% for the platonic sphalerons, or better.

In the following we discuss successively spherically symmetric 
($N=1$), axially symmetric ($N=2,4$)
and platonic ($N=4$) sphalerons.
We construct these excited sphaleron solutions
for the Higgs self-coupling constant $\lambda=0.125$.
and study their dependence on the coupling constant $\alpha$.

\subsection{Spherically symmetric sphalerons}

Fundamental spherically symmetric sphalerons have been studied
before \cite{kkm-dil,Manka}.
As a function of the coupling constant $\alpha$ 
two branches of fundamental sphaleron solutions exist.
The lower branch of dilatonic sphalerons emerges
from the spherically symmetric sphaleron of Weinberg-Salam theory,
and extends up to a maximal value of the coupling constant,
$\alpha_{\rm max}$.
There it merges with the upper branch of fundamental sphaleron solutions,
which extends back to $\alpha=0$.
On the lower branch the dimensionless energy $E$ 
decreases with increasing $\alpha$, 
whereas on the second branch it increases with decreasing $\alpha$ 
and diverges as $\alpha$ tends to zero.
At the same time the value of the dilaton function at the origin
decreases continuously along both branches.

Considering the limit of vanishing $\alpha$ on the upper branch,
we note that the scaled energy $\alpha E$ and
the value of the dilaton function at the origin
$\phi(0)$ approach finite values,
equal to the energy $E$, respectively $\phi(0)$,
of the fundamental spherically symmetric YMD solution \cite{lav}.
Indeed, introducing the scaled variable $\hat{r} = r/\alpha$,
the Higgs field $\hat{\Phi} = \Phi \alpha$ and the gauge field
$\hat{V}_i = V_i \alpha$, one arrives at an equivalent system
of equations,
in which the limit of vanishing Higgs field $\hat{\Phi}$
corresponds to the limit of vanishing $\alpha$ 
($v \rightarrow 0$) on the upper branch
of the original system.
We exhibit the scaled energy $\alpha E$
and the value of the dilaton function at the origin $\phi(0)$
for the fundamental sphalerons
as a function of $\alpha$ in Fig.~1.
 
\begin{figure}[h]
\parbox{\textwidth}{
\centerline{
%\hspace{-2cm}
\mbox{\epsfysize=6.0cm \epsffile{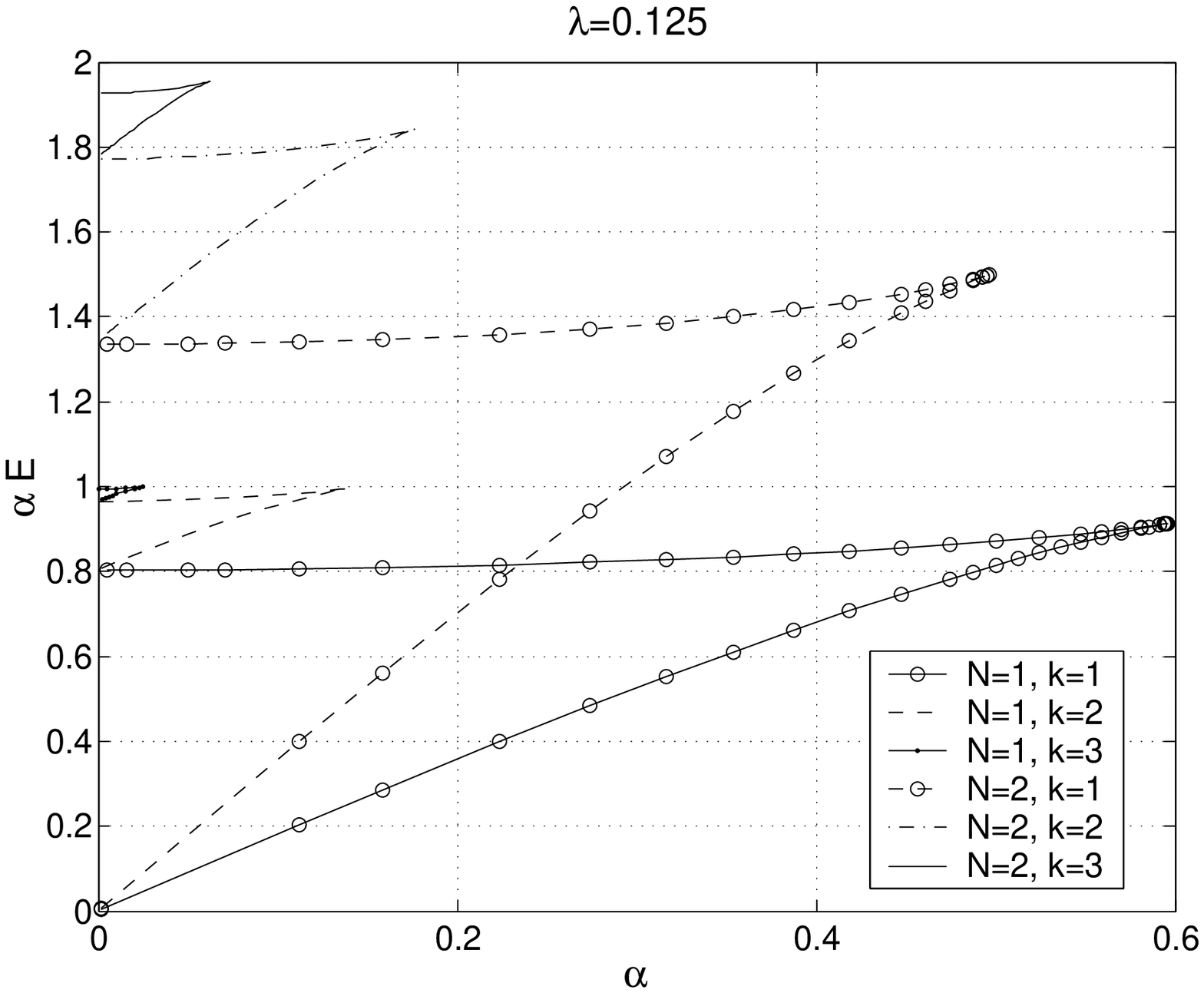} } 
%\hspace{1cm}
\mbox{\epsfysize=6.0cm \epsffile{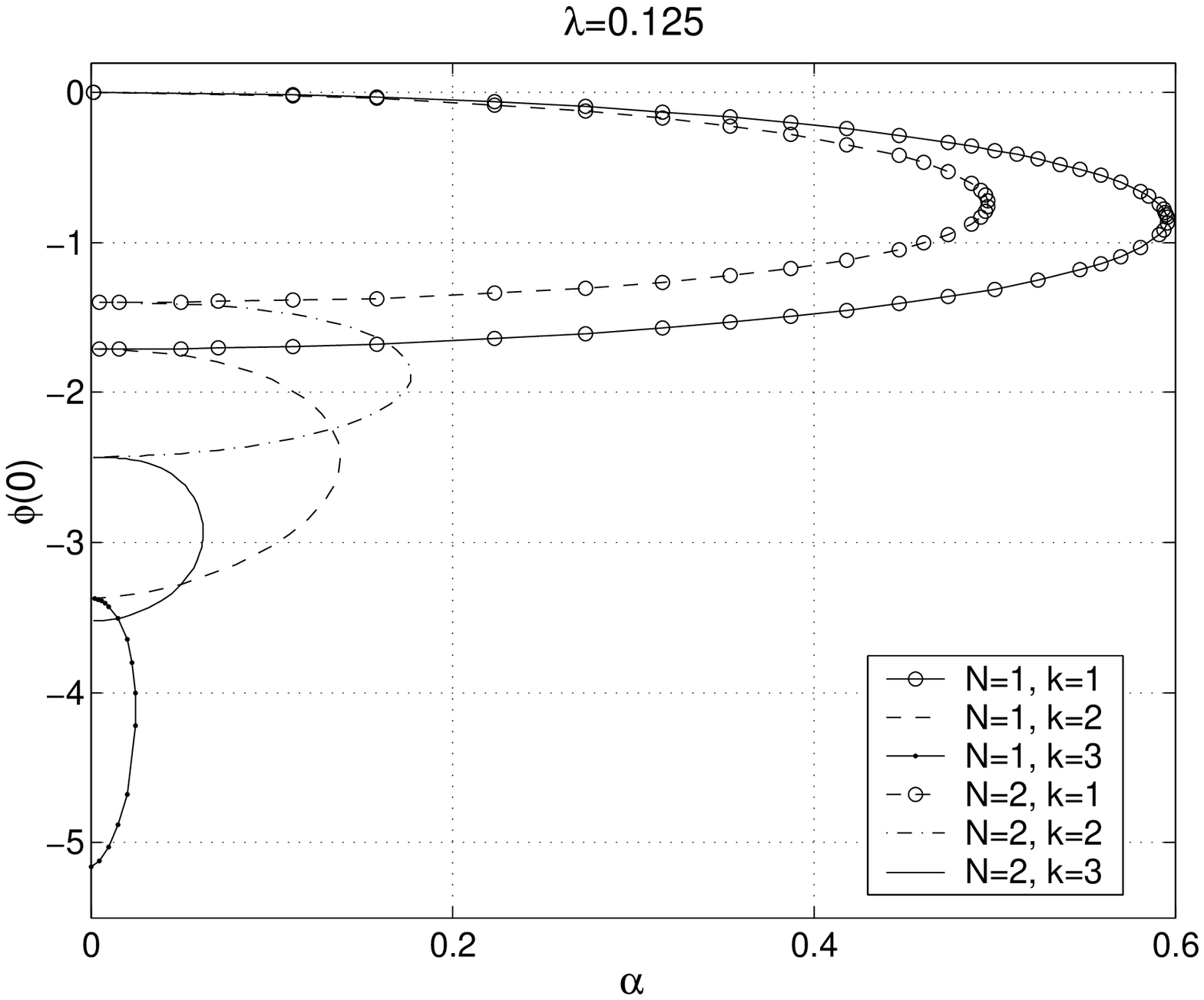} }
}\vspace{0.5cm} 
{\bf Fig.~1} \small
The scaled energy $\alpha E$ (left)
and the value of the dimensionless dilaton field
at the origin $\phi(0)$ (right)
of spherically symmetric sphalerons ($n=1$, $k=1-3$)
and of axially symmetric sphalerons ($n=2$, $k=1-3$)
are shown as functions of the coupling constant $\alpha$
for the Higgs self-coupling constant $\lambda=0.125$.
\vspace{0.5cm}
}
\end{figure}

The figure also exhibits the first three radial excitations
of the spherically symmetric sphalerons, possessing $k=2$, and 3 nodes,
respectively.
For each radial excitation two branches of solutions appear,
which merge and end at a maximal value of the coupling constant
$\alpha_{\rm max}(k)$, which decreases with increasing $k$.
At $\alpha=0$ the scaled energy of the upper branch solutions with $k$ nodes
and the scaled energy of the lower branch solutions with $k+1$ nodes coincide,
yielding for the scaled energy of the solutions
a `Christmas tree' like structure.
For the YMHD solutions themselves the transition
at $\alpha=0$ is not continuous, however.
The scaled YMHD solution with $k$ nodes approaches on its upper branch
smoothly the YMD solution with $k$ nodes \cite{lav}.
But the scaled YMHD solution with $k+1$ nodes emerges 
on its lower branch from the YMD solution with $k$ nodes
with a discontinuity at infinity,
since the additional node of the gauge field function 
has been pushed out to infinity.
Comparing these YMHD solutions with the corresponding solutions of 
EYMH theory \cite{greene}, we observe, that their structure 
and their coupling constant dependence are completely analogous,
in accordance with our expectation.

\subsection{Axially symmetric sphalerons}

For the axially symmetric sphalerons we observe the same pattern
for the fundamental and excited solutions 
as for the spherically symmetric sphalerons.
For a given $n$, the scaled energy of 
the fundamental and excited sphaleron branches 
also exhibits a `Christmas tree' like structure, 
as shown in Fig.~1 for $n=2$.
For $\alpha \rightarrow 0$,
the lower fundamental YMHD sphaleron branches emerge smoothly from the
corresponding sphalerons of Weinberg-Salam theory \cite{kksph}.
The scaled upper YMHD sphaleron branches with $k$ nodes approach
smoothly the corresponding axially symmetric YMD solutions 
with $k$ nodes \cite{kk-dil},
and the lower YMHD sphaleron branches with $k+1$ nodes
emerge with a discontinuity at infinity.
 
\subsection{Cubic sphalerons}

Turning now to the numerical results for the platonic sphalerons
with cubic symmetry,
we again observe the same pattern as for the spherically
and axially symmetric sphalerons.
The scaled energy of
the fundamental and first excited platonic sphaleron branches
exhibits the typical `Christmas tree' like structure, as seen in Fig.~2.
where the value of the dilaton function at the origin
is also shown.

In particular, the lower branch of the fundamental cubic YMHD sphaleron 
emerges smoothly from the cubic sphaleron of Weinberg-Salam theory \cite{kkm},
the scaled upper branch of the fundamental cubic YMHD sphaleron 
approaches the fundamental cubic YMD solution \cite{kkmnew},
and the scaled upper branch of the first excited cubic YMHD sphaleron 
approaches the first excited cubic YMD solution,
giving first evidence for the existence of a whole sequence
of cubic YMD solutions \cite{kkmnew}.

\begin{figure}[h]
\parbox{\textwidth}{
\centerline{
%\hspace{-2cm}
\mbox{\epsfysize=6.0cm \epsffile{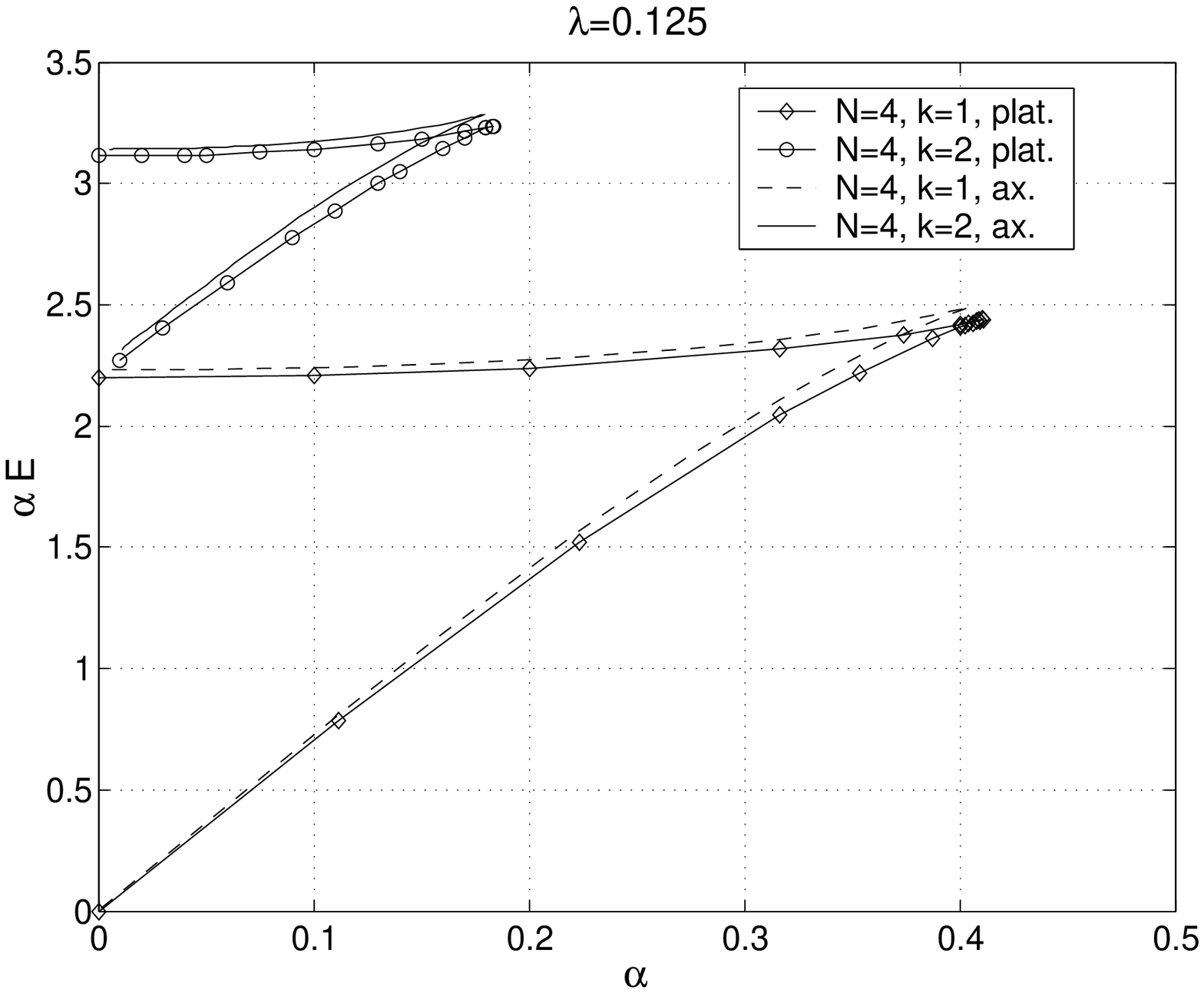} }
%\hspace{1cm}
\mbox{\epsfysize=6.0cm \epsffile{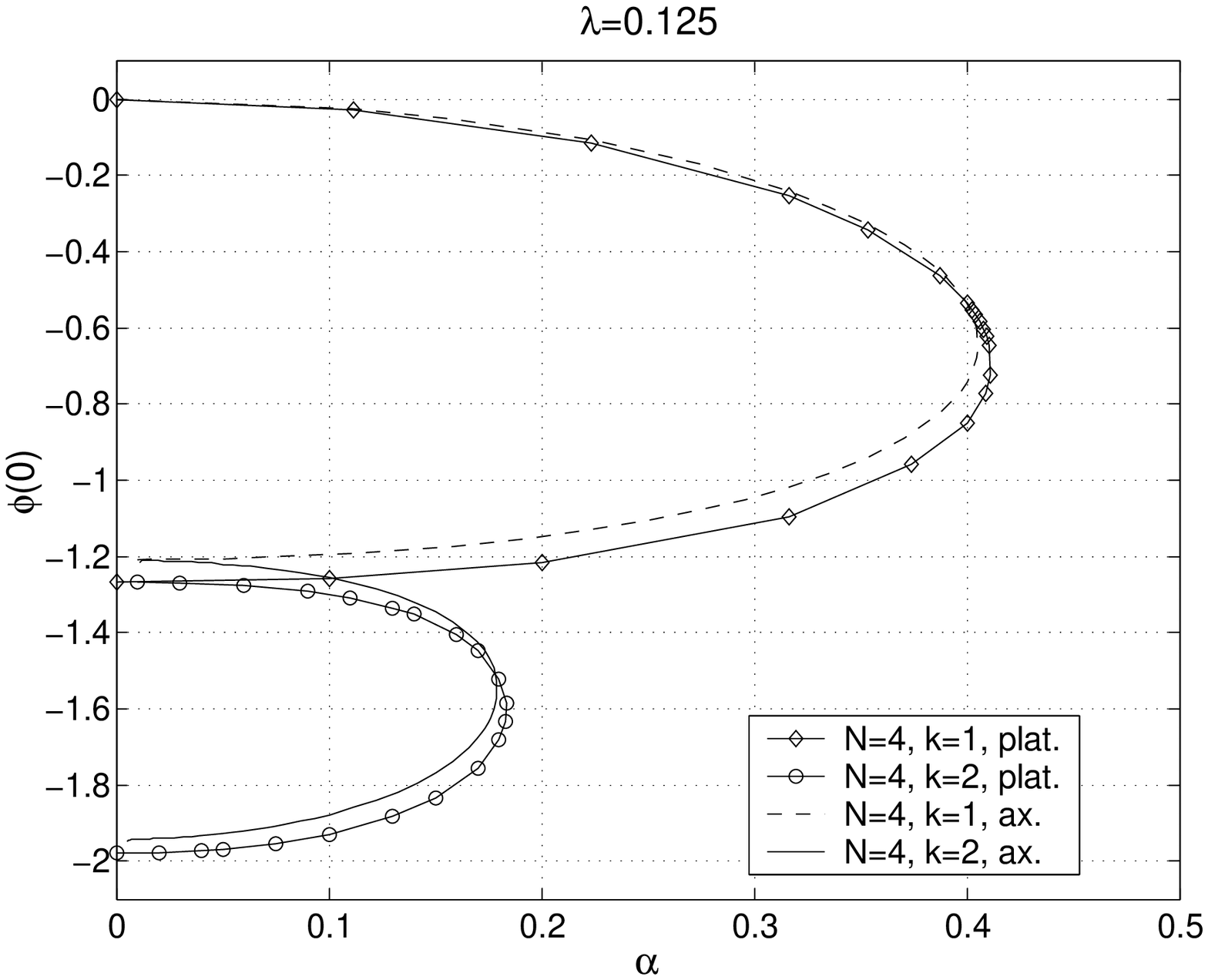} }
}\vspace{0.5cm}
{\bf Fig.~2} \small
The scaled energy $\alpha E$ (left)
and the value of the dimensionless dilaton field
at the origin $\phi(0)$ (right)
of cubic sphalerons ($N=4$, $k=1-2$)
and of axially symmetric sphalerons ($n=4$, $k=1-2$)
are shown as functions of the coupling constant $\alpha$
for the Higgs self-coupling constant $\lambda=0.125$.
\vspace{0.5cm}
}
\end{figure}

Comparison of the platonic sphalerons and the axially symmetric 
sphalerons reveals, that the platonic sphalerons exist 
for slightly larger values of the coupling constant $\alpha$
than the axially symmetric sphalerons.
For all values of $\alpha$, for which both types of
sphalerons coexist, the energy of the platonic sphalerons 
is slightly smaller than the energy of the axially symmetric sphalerons.

Defining the dimensionless energy density $\epsilon$ by
\begin{equation}
E= \int \epsilon (\vec x) dx dy dz
\ , \end{equation}
we present surfaces of constant total energy density $\epsilon_{\rm tot}$ 
in Fig.~3
for the first excited cubic sphaleron 
for $\alpha=0.15$ and $\lambda=0.125$.
Again we conclude 
that the shape of the energy density of platonic solutions
is determined primarily by the rational map \cite{kkm,kkm-dil}.
Interestingly, the energy density of
the excited solutions exhibits a cubic shape both for
large and for small constant values,
revealing a small interior cube within a larger exterior cube.
Surfaces of constant energy density 
of the gauge field $\epsilon_{\rm gauge}$ 
($\frac{1}{2} e^{2\kappa \phi} {\rm Tr} (F_{\mu\nu} F^{\mu\nu})$)
exhibit an analogous pattern,
as also seen in Fig.~3.

%\newpage

\begin{figure}[p]
\parbox{\textwidth}{
\centerline{
\mbox{\epsfysize=6.0cm \epsffile{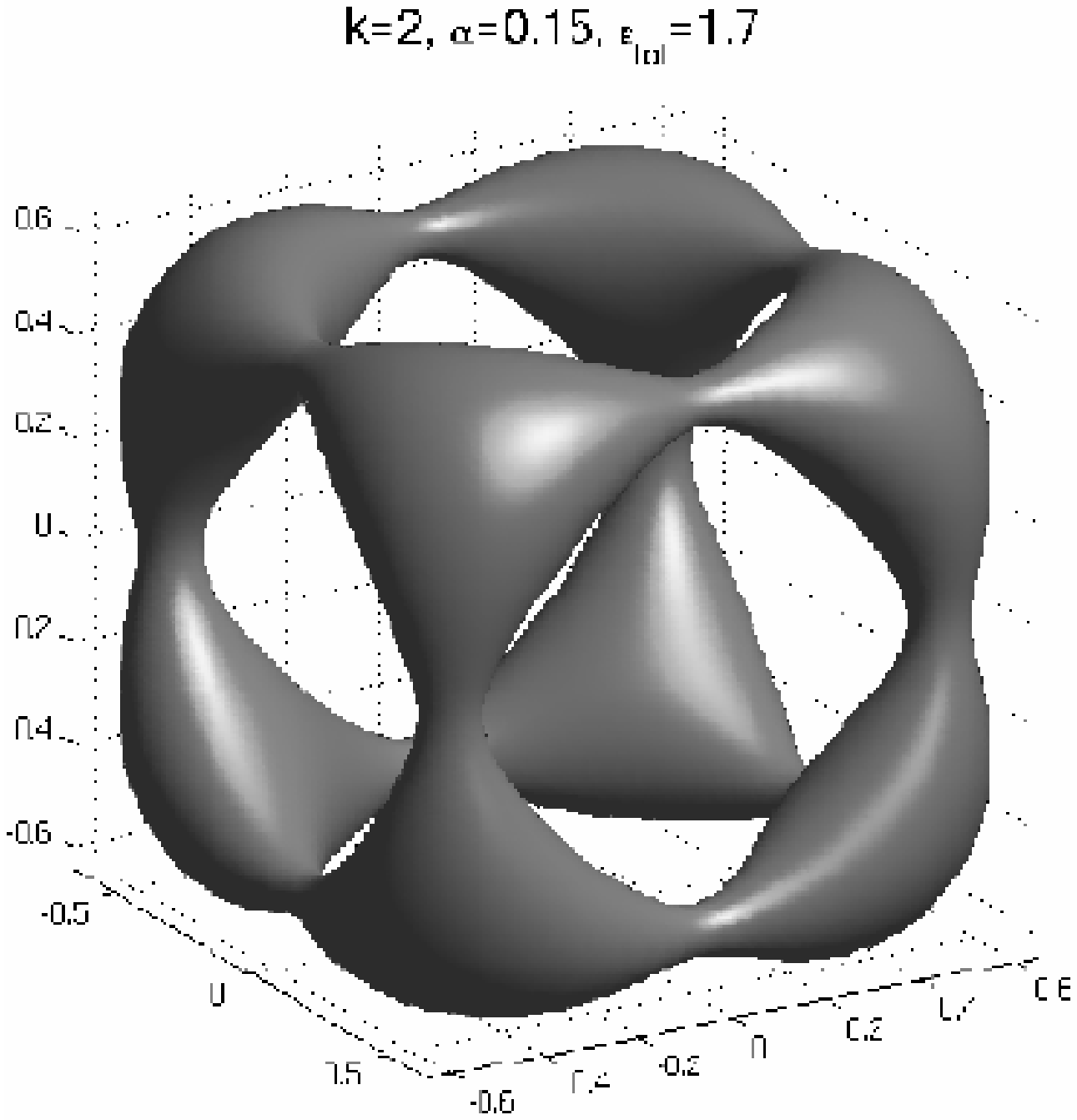} } \hspace{1cm}
\mbox{\epsfysize=6.0cm \epsffile{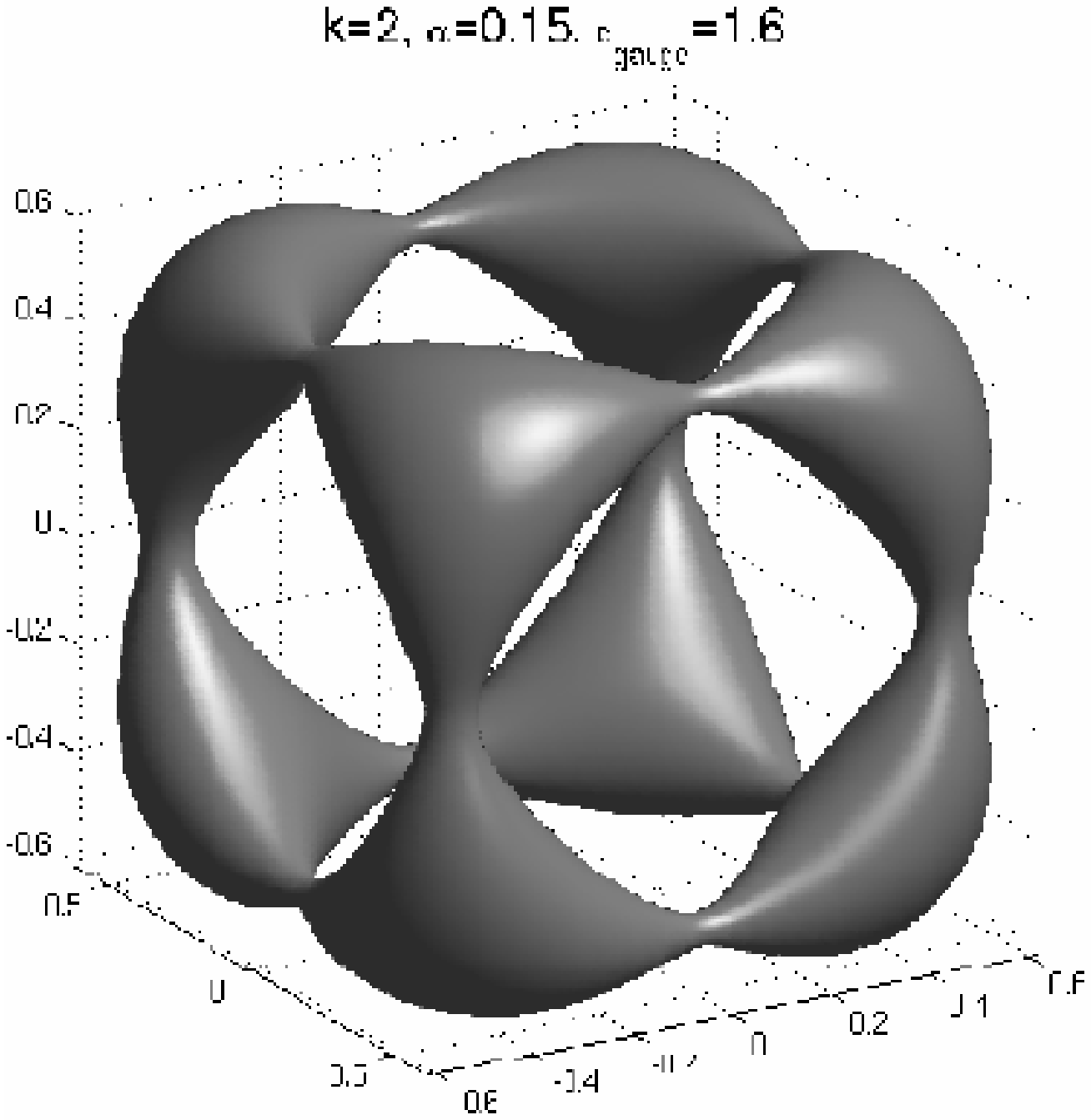} }
}\vspace{1.cm} }
\parbox{\textwidth}{
\centerline{
\mbox{\epsfysize=6.0cm \epsffile{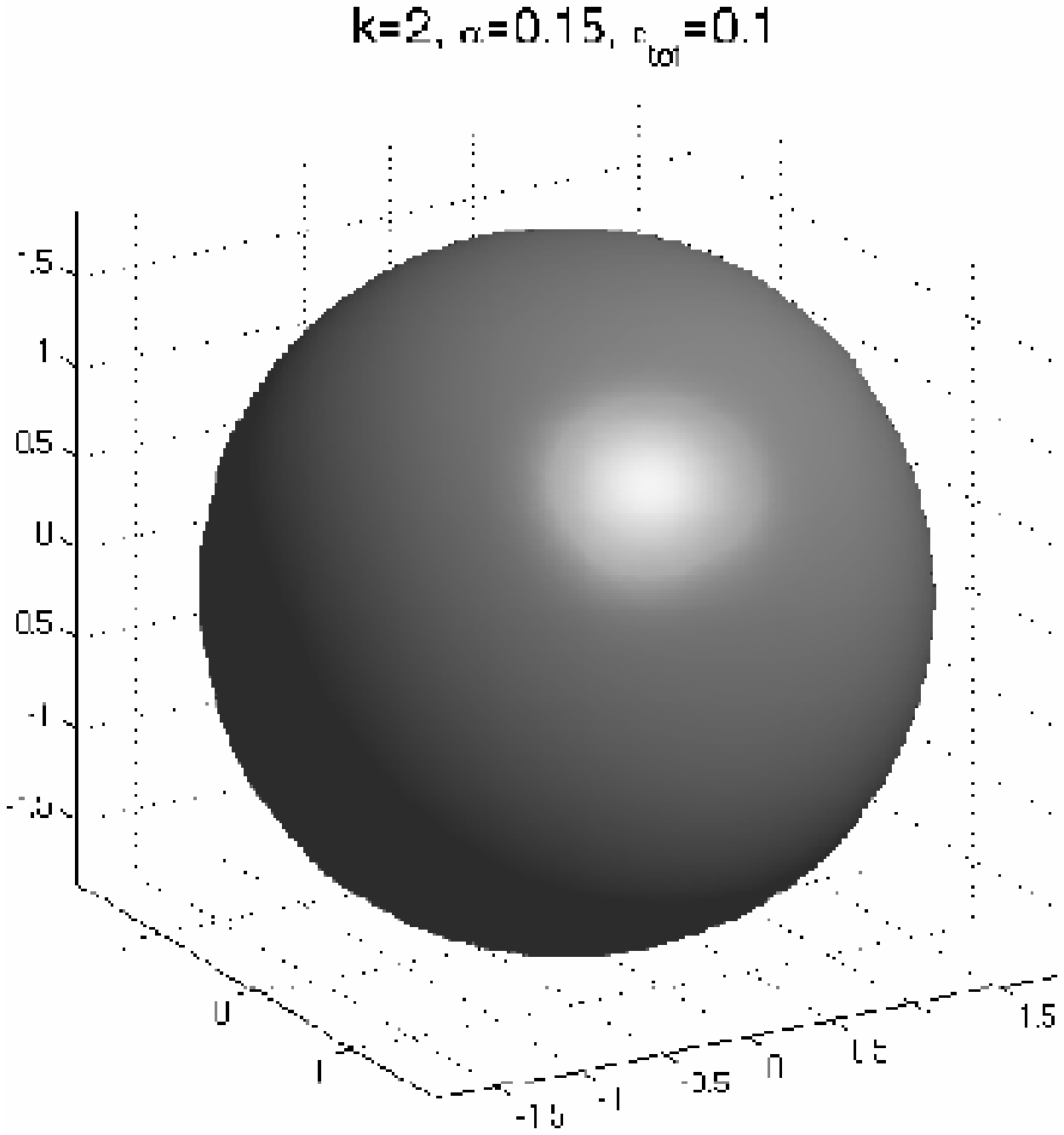} } \hspace{1cm}
\mbox{\epsfysize=6.0cm \epsffile{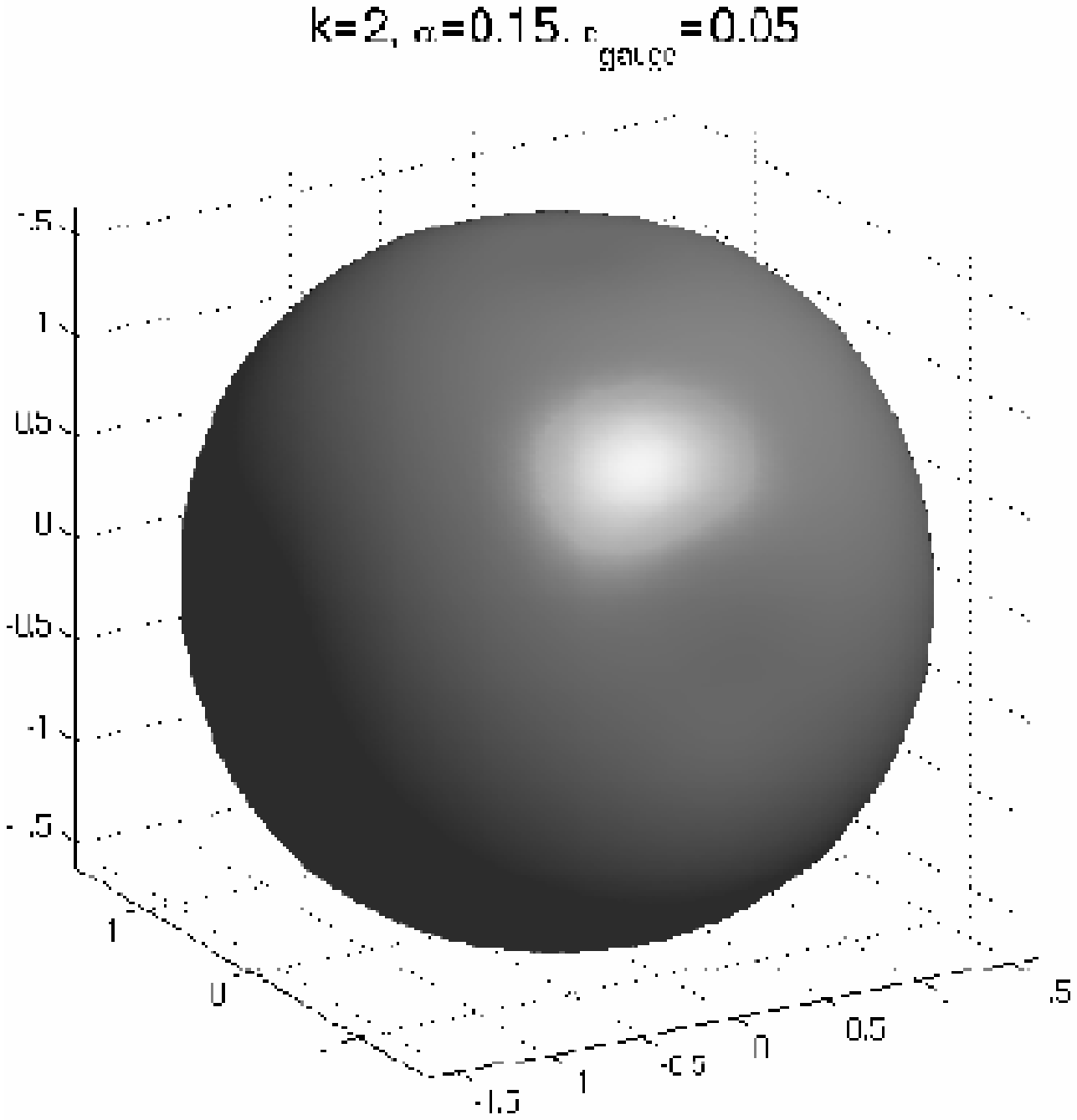} }
}\vspace{1.cm} }
\parbox{\textwidth}{
\centerline{
\mbox{\epsfysize=6.0cm \epsffile{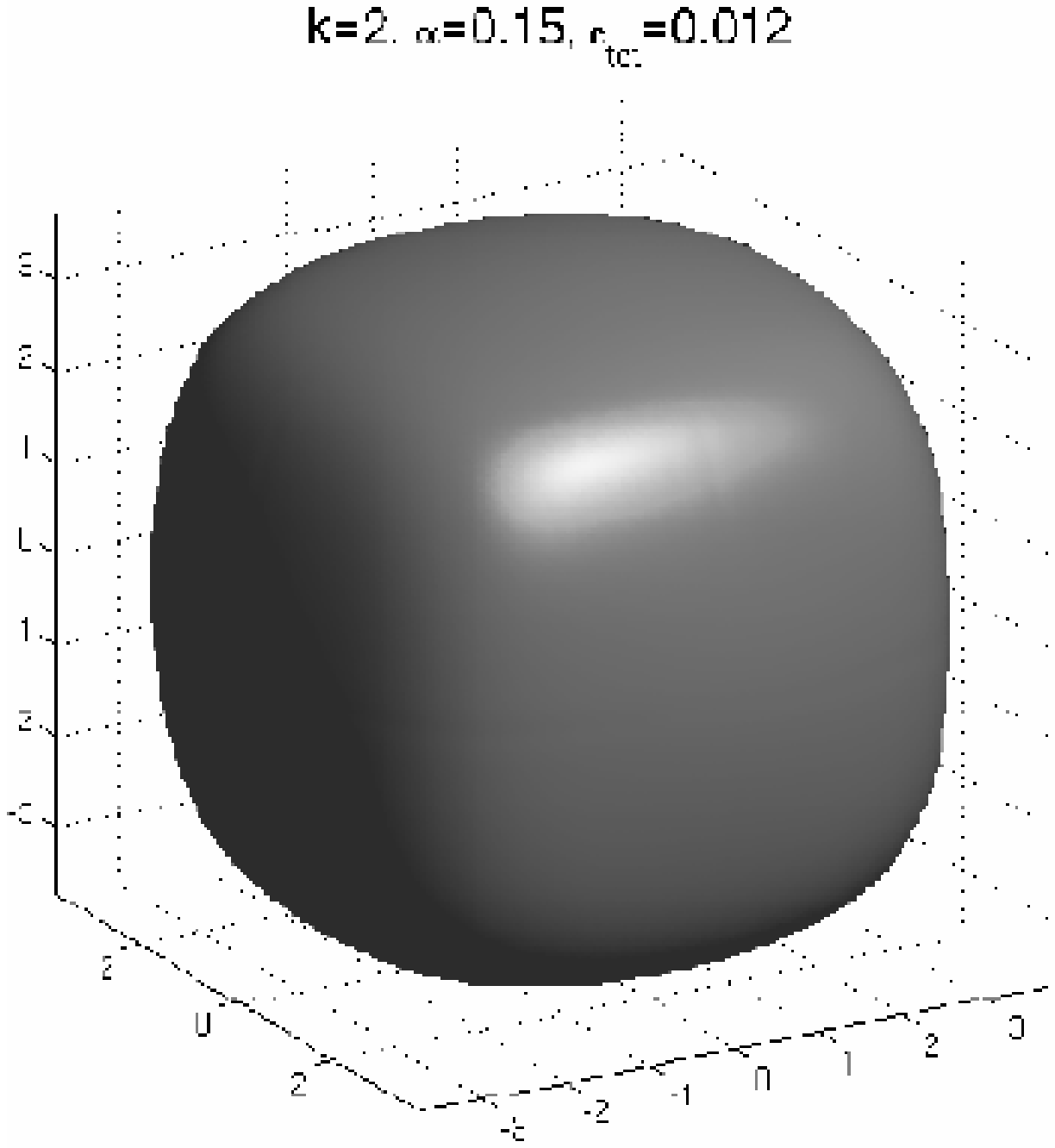} } \hspace{1cm}
\mbox{\epsfysize=6.0cm \epsffile{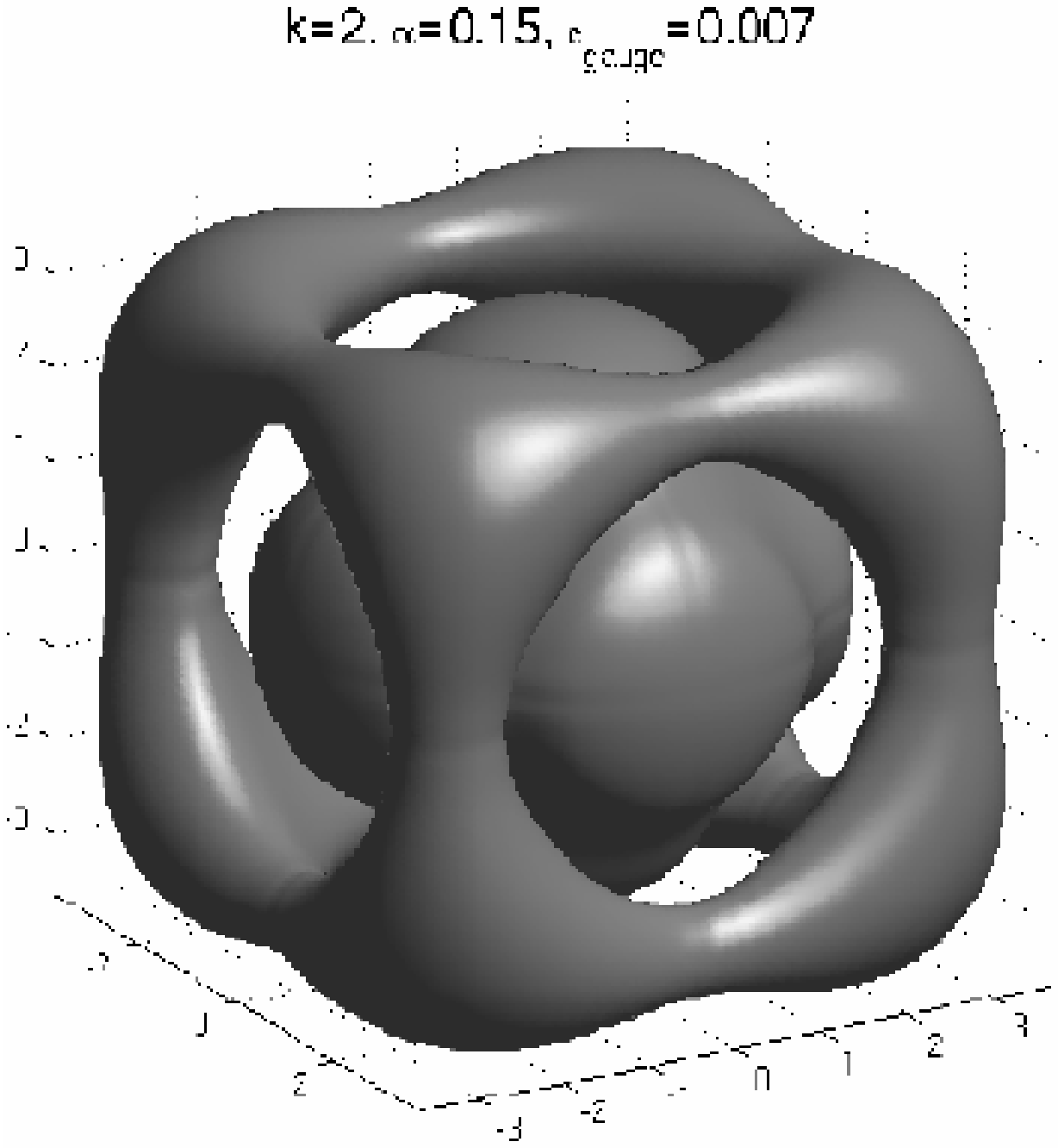} }
}\vspace{1.cm} }
{\bf Fig.~3} \small
Surfaces of constant total energy density $\epsilon$ 
(left column)
and of constant energy density of the gauge field $\epsilon_{\rm gauge}$
(right column)
are shown for the first excitation of the cubic sphaleron
at $\alpha=0.15$ and Higgs self-coupling constant $\lambda=0.125$.
\vspace{0.5cm}
\end{figure}

%\newpage

\section{Conclusions}

We have constructed numerically sphaleron solutions of YMHD theory,
possessing spherical, axial, and cubic symmetry.
We have shown, that to each fundamental sphaleron solution
excited sphaleron solutions exist.
In particular, we have obtained the first excited platonic
solutions.
Previously, only fundamental platonic solutions
were known \cite{monopla,kkm,kkm-dil,plato2}.

For all these types of sphaleron solutions, 
and in particular also for the fundamental and excited sphaleron solutions, 
two branches of solutions exist,
which extend from $\alpha=0$ to a value $\alpha_{\rm max}$,
where they merge and end.
The value of $\alpha_{\rm max}$ depends on the particular type of solution. 
$\alpha_{\rm max}$ decreases with increasing
Chern-Simons number of the solution \cite{kkm-dil}
and also with increasing excitation of the solution.
Thus all solutions reveal the same general coupling constant dependence.

The excited solutions fall into two classes, 
distinguished by their boundary conditions and Chern-Simons numbers.
We observe, that the solutions of both classes alternate
with increasing excitation energy. 
Given a particular fundamental sphaleron solution,
its scaled energy and the energy of its excitations
exhibits a `Christmas tree' like structure.
While we have constructed only the first excited platonic sphalerons,
we conjecture, that higher excitations of platonic solutions exist as well,
analogous to the sequences of excited spherically symmetric
and axially symmetric sphalerons.

In the limit $\alpha \rightarrow 0$, 
the lower fundamental YMHD sphaleron branches 
emerge from the corresponding sphaleron solutions of Weinberg-Salam theory.
All other branches merge into solutions of YMD theory
in the limit $\alpha \rightarrow 0$, after appropriate rescaling.
While the limiting spherically and axially symmetric YMD solutions 
were known before \cite{lav,kk-dil}, 
we have obtained new numerical evidence for the existence 
of platonic YMD solutions \cite{kkmnew}.

We consider the construction of platonic solutions
in the presence of a dilaton 
as a step towards obtaining gravitating
solutions without rotational symmetries. 
By analogy with the simpler spherically symmetric case,
we expect that gravitating platonic sphalerons will
exhibit similar properties as the dilatonic platonic sphalerons
studied here.

{\bf Acknowledgement}: 

B.K.~gratefully acknowledges support by the DFG under contract
KU612/9-1, and K.M.~by the Research Council of Norway under 
contract 153589/432.

%\vfill\eject


\begin{thebibliography}{000}

\bibitem{thooft} 
 G.~`t Hooft,
 Nucl. Phys. {\bf B79} (1974) 276;\\
 A.~M.~Polyakov,
 Pis'ma JETP {\bf 20} (1974) 430.

\bibitem{km} 
 N.~S. Manton,
 Phys. Rev. {\bf D28} (1983) 2019;\\
 F.~R. Klinkhamer, and N.~S. Manton,
 Phys. Rev. {\bf D30} (1984) 2212.

\bibitem{bi}
 J. Kunz and Y. Brihaye,
 Phys. Lett. {\bf 216B} (1989) 353.

\bibitem{yaffe}
 L. Yaffe,
 Phys. Rev. {\bf D40} (1989) 3463.

\bibitem{monoax}
 C.~Rebbi and P.~Rossi,
 Phys. Rev. {\bf D22} (1980) 2010;\\
 R.~S.~Ward,
 Comm. Math. Phys. {\bf 79} (1981) 317;\\
 P.~Forgacs, Z.~Horvath and L.~Palla,
 Phys. Lett. {\bf 99B} (1981) 232;\\
 B. Kleihaus, J. Kunz and D.~H. Tchrakian,
 Mod. Phys. Lett. {\bf A13} (1998) 2523.

\bibitem{kksph} 
 B. Kleihaus, and J. Kunz,
 Phys. Lett. {\bf B329} (1994) 61;\\
 B. Kleihaus, and J. Kunz,
 Phys. Rev. {\bf D50} (1994) 5343.

\bibitem{monopla}
 N.~J. Hitchin, N.~S. Manton and M.~K. Murray,
 Nonlinearity {\bf 8} (1995) 661;\\
 C.~J. Houghton and P.~M. Sutcliffe,
 Commun. Math. Phys. {\bf 180} (1996) 343;\\
 C.~J. Houghton and P.~M. Sutcliffe,
 Nonlinearity {\bf 9} (1996) 385;\\
 P.~M. Sutcliffe,
 Int. J. Mod. Phys. {\bf A 12} (1997) 4663.

\bibitem{kkm} 
 B. Kleihaus, J. Kunz, and K. Myklevoll,
 Phys. Lett. {\bf B582} (2004) 187.

\bibitem{ratmap} 
 C.~J. Houghton, N.~S. Manton and P.~M. Sutcliffe,
 Nucl. Phys. {\bf B 510} (1998) 507.

\bibitem{maison}
 D. Maison,
%Uniqueness of the Prasad-Sommerfield Monopole Solution
 Nucl. Phys. {\bf B182} (1981) 144.

\bibitem{gmono}
 K. Lee, V.P. Nair and E.J. Weinberg,
 Phys. Rev. {\bf D45} (1992) 2751;\\
 P. Breitenlohner, P. Forgacs and D. Maison,
 Nucl. Phys. {\bf B383} (1992) 357;\\
 P. Breitenlohner, P. Forgacs and D. Maison,
 Nucl. Phys. {\bf B442} (1995) 126.

\bibitem{greene}
 B.~R. Greene, S.~D. Mathur, and C.~M. O'Neill,
 Phys. Rev. {\bf D47} (1993) 2242;\\
 Y. Brihaye, and M. Desoil,
 Mod. Phys. Lett. {\bf A15} (2000) 889.

\bibitem{bm}
 R. Bartnik and J. McKinnon,
 Phys. Rev. Lett. {\bf 61} (1988) 141.

\bibitem{vg}
 M.~S. Volkov and D.~V. Gal'tsov,
 Phys. Rept. {\bf 319} (1999) 1.

\bibitem{lav}
 G. Lavrelashvili and D. Maison,
 Phys. Lett. {\bf B295} (1992) 67;\\
 P. Bizon,
 Phys. Rev. {\bf D47} (1993) 1656;\\
 D. Maison, 
 Commun. Math. Phys. {\bf 258} (2005) 657.

\bibitem{forgacs}
 P. Forgacs and J. Gy\"ur\"usi,
 Phys. Lett. {\bf B366} (1996) 205.

\bibitem{kk}
 B. Kleihaus and J. Kunz,
 Phys. Rev. Lett. {\bf 78} (1997) 2527;\\
 B. Kleihaus and J. Kunz,
 Phys. Rev. {\bf D57} (1998) 834.

\bibitem{kk-dil}
 B. Kleihaus and J. Kunz,
 Phys. Lett. {\bf B392} (1997) 135.

\bibitem{kkm-dil}
 B. Kleihaus, J. Kunz and K. Myklevoll,
%Platonic Sphalerons in the Presence of a Dilaton Field,
 Phys. Lett. {\bf B605} (2005) 151.

\bibitem{kkb} 
 B. Kleihaus, J. Kunz, and Y. Brihaye,
 %The electroweak sphaleron at physical mixing angle,
 Phys. Lett. {\bf B273} (1991) 100;\\
 J. Kunz, B. Kleihaus, and Y. Brihaye,
 %Sphalerons at finite mixing angle,
 Phys. Rev. {\bf D46} (1992) 3587.

\bibitem{kb} 
 Y. Brihaye and J. Kunz,
 %Axially Symmetric Solutions in the Electroweak Theory,
 Phys. Rev. {\bf D50} (1994) 4175.

\bibitem{fidisol}
 W.~Sch\"onauer and R.~Wei\ss,
 J.~Comput. Appl. Math, {\bf 27} (1989) 279;\\
 M.~Schauder, R.~Wei\ss \ and W.~Sch\"onauer,
 The CADSOL Program Package, Universit\"at Karlsruhe,
 Interner Bericht Nr. 46/92 (1992).

\bibitem{Manka}
 D.~Karczewska and R.~Manka,
 arXiv:hep-th/9612020;
 Phys.\ Scripta {\bf 63} (2001) 87.

\bibitem{kkmnew}
 B. Kleihaus, J. Kunz, and K. Myklevoll,
 in preparation.

\bibitem{plato2} 
 E. Braaten, S. Townsend and L. Carson,
 Phys. Lett. {\bf B235} (1990) 147;\\
 R.~A. Battye and P.~M. Sutcliffe,
 Phys. Rev. Lett. {\bf 79} (1997) 363;\\
 R.~A. Battye and P.~M. Sutcliffe,
 Phys. Lett. {\bf B416} (1998) 385;\\
 D.Yu. Grigoriev, P.M. Sutcliffe, D.H. Tchrakian,
 Phys. Lett. {\bf B540} (2002) 146.

%\bibitem{}

\end{thebibliography}
\end{document}